\documentclass[preprint,aps,prd,
nofootinbib,superscriptaddress,
tightenlines]{revtex4}
\usepackage{bm}

\usepackage{float}
\usepackage{dcolumn}
\usepackage{graphics}
\usepackage{amsmath, amsfonts, amsthm, amssymb, accents, graphicx,comment}
\usepackage{etex}
\usepackage{hyperref,subfigure}
\usepackage{footnote}

\newcommand{\n}{\nonumber}
\newcommand{\bea}{\begin{eqnarray}}
\newcommand{\eea}{\end{eqnarray}}

\allowdisplaybreaks[3]
\def \bal#1\eal  {\begin{align} #1 \end{align}}
\newcommand{\be} {\begin{equation}}
\newcommand{\ee} {\end{equation}}
\newcommand{\bpm}{\begin{pmatrix}}
\newcommand{\epm}{\end{pmatrix}}
\newcommand{\nn} {\nonumber\\}

\newcommand{\ai}{{\alpha}}

\newcommand{\ri}{{\rho}}
\newcommand{\si}{{\sigma}}

\begin{document}

\title{Relativistic star solutions in Mass-varying Massive Gravity with a diagonal metric}

\author{De-Jun Wu}
\email[]{wudejun10@mails.ucas.ac.cn}
\affiliation{School of Science, Inner Mongolia University of Science and Technology, Inner Mongolia, Baotou 014010, China}

\begin{abstract}
We investigate relativistic star solutions in Mass-Varying Massive Gravity (MVMG) with a diagonal metric. Contrary to the intuition that there is no fundamental difference between diagonal metric and non-diagonal metric solutions regarding relativistic stars, we find that with a diagonal metric, well-behaved relativistic star solutions may not exist except for trivial ones in which the graviton mass is a constant, whereas non-trivial relativistic star solutions had been found in MVMG with a non-diagonal metric. The reason is that with a diagonal metric, the field equations constitute a system of differential-algebraic equations of differential index-2 with two extra constraints that have a significant influence on the system, rendering the relativistic star solution with a non-trivial graviton mass configuration impossible in most cases. 
\end{abstract}

\maketitle
\section{Introduction}
To develop a self-consistent theory of gravity with a non-zero mass was a historical effort, popularized through late developments \cite{deRham:2010ik,deRham:2010kj,Hassan:2011hr,Hassan:2011tf,Hassan:2011ea,deRham:2011rn}.
After overcoming some difficulties,
the final theory takes a standard form known as dRGT which adds a potential term to General Relativity (GR) \cite{deRham:2014zqa,Hinterbichler:2011tt}. 
Several variations and extensions of dRGT have also been developed. In bigravity, the reference metric is promoted to be dynamical \cite{Hassan:2011zd}. A new scalar degree of freedom is introduced in both quasi-dilaton \cite{DAmico:2012hia,DeFelice:2013tsa} and Mass-Varying Massive Gravity (MVMG)  \cite{Huang:2012pe,Huang:2013mha}. MVMG would enable the interesting phenomenon that graviton mass could become extremely
large near massive objects such as black holes, yet small enough on large scale to satisfy cosmological observations \cite{Zhang:2017jze}.

In addition to providing a theoretical framework to explain the late-time cosmic acceleration \cite{Riess:1998cb,Perlmutter:1998np}, like many other theories of modified gravity \cite{Clifton:2011jh}, Massive Gravity may also play a role in understanding some results of the astrophysics observations, especially due to recent developments in gravitational waves detection \cite{Abbott:2016blz,TheLIGOScientific:2017qsa}. For example, new gravitational wave polarizations exist in Massive Gravity because some degrees of freedom that GR eliminated are reintroduced. Massive Gravity also has a different dispersion relation from GR which will have observable effects \cite{deRham:2016nuf}. Besides directly influencing the way gravitational waves propagate, Massive Gravity may also change the processes gravitational waves emit, which usually involve massive astronomical objects such as neutron stars and black holes   \cite{TheLIGOScientific:2017qsa}. 

When studying static, spherically symmetric compact star solutions in GR, whether the physical metric is diagonal or not is not a matter of concern due to the presence of the diffeomorphism invariance.  However,  the differences between diagonal and non-diagonal metric solutions in Massive Gravity are not inconsequential since the diffeomorphism invariance is explicitly broken. With a non-diagonal physical metric, black hole solutions have been found in MVMG \cite{Tolley:2015ywa,Zhang:2017jze}, as well as relativistic star solutions \cite{Sun:2019niy}. 
With a diagonal physical metric, relativistic star solutions have been found in the minimal coupled dRGT model \cite{Katsuragawa:2015lbl,Yamazaki:2017ztk}. However, a similar study trying to find relativistic star solutions in the full dRGT model was met with some difficulties \cite{Yamazaki:2018szv}. In the studies of black hole solutions in dRGT or MVMG, a diagonal physical metric is generally avoided due to the problems associated with the singularities at horizons and a non-diagonal physical metric is preferable \cite{Berezhiani:2011mt, Babichev:2015xha}. It is only considered when the reference metric is no longer assumed to be flat. Should such conclusions be extended to compact star solutions in MVMG?  Do new difficulties arise when finding solutions with a diagonal physical metric? We try to answer the questions by investigating the static, spherically symmetric relativistic star solutions with a  diagonal physical metric in MVMG. It also serves as a case study of the differences between non-diagonal and diagonal solutions in Massive Gravity.

Relativistic star solutions in Massive Gravity with a diagonal physical metric and a singular reference metric have also been investigated \cite{Hendi:2017ibm, EslamPanah:2018evk}. A singular reference metric would cause the theory to deviate from standard dRGT, such theories are not the focus of this study. 
The paper is organized as follows, in Section \ref{sec:setup} we discuss briefly the MVMG model and the setup. In Section \ref{sec:field} the field equations in the minimal model are given and we will investigate their unusual nature thoroughly. The field equations turn out to be a system of differential-algebraic equations (DAEs) with constraints. We study the effects of the extra constraints on the system in Section \ref{sec:branching} and possible asymptotically flat solutions in Section \ref{sec:asymptotically flat solution}. Both the minimal and full MVMG models are considered.
Finally, we summarize all the findings in Section \ref{sec:summary}. We shall use geometric units which set $c = G = 1$.

\section{MASS-VARYING Massive Gravity model and Setup}
\label{sec:setup}

The action of Mass-varying Massive Gravity reads
\be
S = \frac{1}{  8\pi} \int {\text{d}}^4 x \sqrt{-g}\bigg[\frac{\cal R}{2}+V(\sigma) U(\mathcal{K})
-\frac{1}{2} g^{\mu \nu} \partial_{\mu} \sigma \partial_{\nu} \sigma-W(\sigma)\bigg]+S_{\text{m}}\left[g_{\mu \nu}\right],
\ee
$U(\mathcal{K})$ is the graviton potential and it is given by 
\be
U(\mathcal{K}) =  \mathcal{K}_{\left[\mu\right.}^{\mu} \mathcal{K}_{\left.\nu\right]}^{\nu}+\ai_3 \mathcal{K}_{\left[\mu\right.}^{\mu} \mathcal{K}_{\nu}^{\nu} \mathcal{K}_{\left.\ri\right]}^{\ri} + \ai_4 \mathcal{K}_{\left[\mu\right.}^{\mu} \mathcal{K}_{\nu}^{\nu}\mathcal{K}_{\ri}^{\ri} \mathcal{K}_{\left.\si\right]}^{\si} ,
\ee
with
$\mathcal{K}_{\nu}^{\mu} \equiv \delta_{\nu}^{\mu}-\sqrt{g^{\mu\ri} f_{\ri\nu} }$, $f_{\ri\nu}$ is the reference metric, $\alpha_3$ and $\alpha_4$ are free parameters.   $S_{\text{m}}$ is the matter action.

We consider the ansatz in which both the physical and the reference metric are static, spherically symmetric and diagonal,
\begin{align}
\mathrm{d} s^{2}&=-e^{2a(r)} \mathrm{d}t^{2}+e^{2b(r)} \mathrm{d}r^{2}+r^2 \mathrm{d} \Omega^{2}, \\    
{\mathrm{d} s^{2}}_f&=- \mathrm{d}t^{2}+ d'(r)^2 \mathrm{d}r^{2}+d(r)^2 \mathrm{d} \Omega^{2} .
\end{align}
The setup is equivalent to a diagonal physical metric with an unknown radial component and a Minkowski reference metric.
The scalar field $\sigma \equiv \sigma(r)$ is also static and spherically symmetric. 
We set the mass potential and the scalar potential as 
\bea
\label{vform}
V\left( \sigma  \right) &=& m + {V_\sigma }{\sigma ^n},\\
\label{wform}
W\left( \sigma  \right) &=& {W_\sigma }{\sigma ^l},
\eea
where ${V_\sigma }$ and $W_\sigma$ are positive, $m$ is non-negative, and $n$, $l$ are even positive integers. Such a choice of the potentials is a good candidate for the approximation of general potentials around a local minimum, it is very common among previous studies in MVMG \cite{Tolley:2015ywa,
Zhang:2017jze,Sun:2019niy}. The value of $m$, i.e., the minimum value of the graviton mass, is at least bounded by the Hubble scale and was sometimes neglected \cite{Zhang:2017jze,Sun:2019niy}.
However,  we find that with a diagonal metric, whether the graviton mass potential is non-vanishing or not makes a fundamental difference in  the system, so we intend to study both cases separately.

The Einstein field equations and the scalar field equation are
\begin{align}
\label{EF}
G_{\mu \nu} &=8 \pi   T_{\mu \nu}+T^{(\si)}_{\mu \nu}+V(\sigma) X_{\mu \nu}, \\
\partial_{\mu}\left(\sqrt{-g} g^{\mu \nu} \partial_{\nu} \sigma\right)&=\sqrt{-g}\left(\frac{\partial W}{\partial \sigma}-\frac{\partial V}{\partial \sigma} U\right),
\end{align}
where $T^{(\si)}_{\mu \nu}$ is the scalar energy-momentum tensor and $X_{\mu \nu}$ the effective energy-momentum tensor, which is given by
\begin{align}
X_{\mu\nu}  = &-({\cal K}_{\mu\nu}  - [{\cal K}] g_{\mu\nu})  + \alpha \left( {{\cal K}_{\mu\nu}^2  - [{\cal K}] {\cal K}_{\mu\nu}  + U_2 g_{\mu\nu} } \right)
 \nn
&- \beta \left( {{\cal K}_{\mu\nu}^3  - [{\cal K}] {\cal K}_{\mu\nu}^2  + U_2 {\cal K}_{\mu\nu}  - U_3 g_{\mu\nu} } \right),    
\end{align}
where $[{\cal K}]$ is the trace of $\cal K$, and
\be
U_2  = {\cal K}_{[\mu }^\mu  {\cal K}_{\nu ]}^\nu  , ~~~U_3  = {\cal K}_{[\mu }^\mu  {\cal K}_\nu ^\nu  {\cal K}_{\rho ]}^\rho . 
\ee
$\alpha$ and $\beta$ are defined as $\alpha \equiv 1 + \alpha_3$ and $\beta \equiv \alpha_3 + \alpha_4$. 
In the following section, we mainly focus on the minimal model as a special case, where the parameters in graviton potential are set as
\be
\alpha = \beta = 0.
\ee
The full model, in which $\alpha$ and $\beta$ can take any value, will be considered in later sections.

\section{Field equations in the minimal model}
\label{sec:field}
In this section, we derive the field equations which turn out to be DAEs and try to transform them into ODEs. 
Two additional constraints emerge in this process and they have a significant impact on the system. We use the minimal model for convenience as the field equations in the full model are rather cumbersome, but the key results still hold. 

The ${^t_t}$, ${^r_r}$ and ${^\theta_\theta}$ components of the Einstein field equations~(\ref{EF}) read
\be
\label{bc1}
 - \frac{{{e^{ - 2b}}\left( {2rb' + {e^{2b}} - 1} \right)}}{{{r^2}}} = \left( {3 - {e^{ - b}}d' - \frac{{2d}}{r} } \right)V - \left( {\frac{1}{2}{e^{ - 2b}}{{\sigma '}^2} + W} \right) - 8\pi \rho,
\ee
\be
\label{con1}
\frac{{{e^{ - 2b}}\left( {2ra' + 1} \right) - 1}}{{{r^2}}} = \left( {3 - {e^{ - a}} - \frac{{2d}}{r} } \right)V + \left( {\frac{1}{2}{e^{ - 2b}}{{\sigma '}^2} - W } \right) + 8\pi p,
\ee
\be
\label{a}
\frac{{{e^{ - 2b}}\left( {\left( {ra' + 1} \right)\left( {a' - b'} \right) + ra''} \right)}}{r} = \left( { 3- {e^{ - a}} - {e^{ - b}}d' - \frac{d}{r}} \right)V - \left( {  \frac{1}{2}{e^{ - 2b}}{{\sigma '}^2} + W} \right) + 8\pi p.
\ee
The field equation of the scalar fields is
\be
\label{s}
{e^{ - 2b}}\sigma ''+\frac{{{e^{ - 2b}}\left( {ra' - rb' + 2} \right)\sigma '}}{r}  = W' - \left( {\frac{{{e^{ - a - b}}{d^2}d'}}{{{r^2}}} - {e^{ - a}} - {e^{ - b}}d' - \frac{{2d}}{r} + 3} \right)V',
\ee
and lastly, there is the equation of energy-momentum conservation,
\be
\label{p}
a'(p + \rho ) + p' = 0.
\ee
The derivatives of mass potential $V$ and scalar potential $W$ are with respect to $\sigma$, the rests are with respect to radius $r$.

Before solving the equations, we shall verify that they do not constitute a system of ODEs in which the derivatives of all unknown functions should in principle be obtainable through algebraic operations. It can be shown that $b'$ and $d'$ cannot be determined algebraically from the system. Eqs.~(\ref{a}), (\ref{s}) and (\ref{p})  can determine $a''$, $\sigma''$ and $p'$. Among the last two equations, (\ref{bc1}) contains $b'$ and $d'$ but (\ref{con1}) does not, making the task of solving  $b'$ and $d'$ algebraically impossible. 
This fact can be made more evident by substituting $\sigma '$ and $a '$ in Eq.~(\ref{con1}) with two new variables, $\Sigma$ and $A$, after which the equation is purely algebraic.
Therefore, these equations do not qualify as a system of ODEs but rather DAEs, with Eq.~(\ref{con1}) as an algebraic constraint.

To transform the system into ODEs that could be solved more easily, we take the derivative of Eq.~(\ref{con1}). The aim is to obtain another differential equation so that $b'$ and $d'$ can be determined explicitly. After some calculations we have
\be
\label{con2}
r{e^a}\left( {ra' - 2{e^b} + 2} \right)V + \sigma '\left( {{r^2}{e^a} - {d^2}} \right)V'=0.
\ee
If $\sigma '$ or $V'$ is set to $0$, in which case dRGT is recovered, Eq.~(\ref{con2}) will reduce to the similar equation obtained in \cite{Katsuragawa:2015lbl} using the condition of conservation of the effective energy-momentum tensor. However, the equation still does not contain $b'$ or $d'$, which is insufficient for the system to transform into ODEs. Therefore, we take the derivative again and obtain
\bea
\label{bc2}
&\frac{1}{2}{e^{a}}\left( {r\left( {2ra'b' - 4{e^b}\left( {a' + b'} \right) + 6a' + 2b' + 2r{e^{2b}}(8\pi p - W) - r{{\sigma '}^2}} \right) - 4{e^b} + 4} \right)V\n\\
& + r{e^b}\left( {{e^a}\left( {{e^b}(3r - d) - rd'} \right) - r{e^b}} \right){V^2} + \left( {{r^2}{e^a} - {d^2}} \right){{\sigma '}^2}V''
\n\\
& + \frac{1}{r}\left( {\sigma '\left( {{d^2}\left( {ra' - rb' + 2} \right) + {r^2}{e^a}\left( {r\left( {a' + b'} \right) - 2{e^b} + 2} \right) - 2rdd'} \right) + r{e^{2b}}\left( {{r^2}{e^a} - {d^2}} \right)W'} \right)V'\n\\
& - \frac{1}{{{r^2}}}{e^{b - a}}\left( {{r^2}{e^a} - {d^2}} \right)\left( {r{e^a}\left( {{e^b}(3r - 2d) - rd'} \right) - {r^2}{e^b} + {d^2}d'} \right){{V'}^2}=0.
\eea
This equation contains $b'$ and $d'$ so their expressions can be determined, but the expressions are too lengthy to write down explicitly. We give two symbolic expressions instead,
\bea
\label{b}
b' &=& B(a', {\sigma }',  a, \sigma, b, d, \rho, r),\\
\label{d}
d' &=& D(a', {\sigma }',  a, \sigma, b, d, \rho, r).
\eea
We transform the system into  ODEs by taking the second order derivative of one existing equation, so that the differential index of the system is 2. We further check if a new constraint would arise from Eq.~(\ref{bc2}) by taking the derivative again and the result is an identity. The above calculations are performed under the assumption that $\sigma '$  is not equal to $0$ across the whole space, otherwise the system reduces to dRGT.

Besides one equation for each physical quantity, Eq.~(\ref{con1}) and Eq.~(\ref{con2}) remain as two additional purely algebraic constraints. In principle, we could use these constraints to obtain an effective Tolman–Oppenheimer–Volkoff (TOV) equation by expressing $b$, $d$ and their derivatives in terms of $a$, $\sigma$, $\sigma'$, $a'$ and $r$.
We can solve $b$ from Eq.~(\ref{con2}) as
\be
b = \ln{\frac{1}{2} \left(r a'+\frac{e^{-a} \left(r^2 e^{a}-d^2\right) \sigma ' V'}{r V}+2\right)}.
\ee
However, when trying to solve $d$ by plugging the above equation into Eq.~(\ref{con1}), we encounter a polynomial equation of degree 5 which is incapable of yielding any algebraic solution. This fact compels us to solve the system in its current form without an effective TOV equation. In the full dRGT model, similar complications may occur even without the scalar field \cite{Yamazaki:2018szv}.

\section{The possibility of branching and the absence of oscillating solutions}
\label{sec:branching}

The fact that $d$ potentially has multiple values indicates that there might be multiple branches of solutions in the system, however, it is not necessarily the case. One can draw a comparison between the system under study and a system of a pendulum moving in a plane, which is a well-known example of DAEs. The pendulum has a constraint regarding its position which reads $x^2+y^2=0$. The fact that the constraint has two solutions does not imply that there are two branches of solutions. Even if the formalism of branching is employed, solutions from different branches can be pieced together smoothly. In other words, branching is not a physical property but rather the result of algebraic manipulations in the case of a pendulum.

We can determine whether relativistic star solutions in MVMG have multiple branches by applying the regularity requirement and finding boundary conditions at the center of the star. If only one approximate solution can be found, yielding one unique boundary condition, then there is no branching in the system. The requirement reads $a' (0) = b' (0) = \sigma' (0) =0$, another condition must be included which is $d(0)=0$, otherwise the system does not have any self-consistent solution. We plug the conditions into the system and can only find one solution. It indicates that there is no branching in relativistic star solutions, and this conclusion holds in both the minimal model and the full model.

However, in some instances there are multiple branches of asymptotic solutions, which will be demonstrated in later sections. This contradiction may lead to some issues in the system. It indicates a mismatch between the parameter space of the solutions at the center of the star and that of infinity. If we assume that different branches correspond to the same approximate solution at the center of the star, there would still be a problem of discontinuity in the limit of vanishing graviton mass, which can be shown clearly through the comparison with compact star solutions of MVMG with a non-diagonal metric.

Branching also occurs in MVMG systems with a non-diagonal metric \cite{Tolley:2015ywa,Zhang:2017jze,Sun:2019niy}, but the key difference is that in such models, the two branches of solutions differ only in the value of a non-dynamical factor while described by the same set of field equations. Their approximate solutions at any point in space are also different which is the result of the difference in said non-dynamical factor;  there is no possibility of mismatch at all.
In the limit of vanishing graviton mass, two branches survive while being consistent with GR. 
In the case of a diagonal metric, if we assume that different branches correspond to the same approximate solution at the center of the star, that one approximate solution would presumably reduce to one GR solution in the limit of vanishing graviton mass, forcing different branches to approach the GR solution. Thus, the number of solutions changes and the limit is not smooth.  The only other possible bypass is that among all the branches of solutions, only one is physical and corresponds to the unique approximate solution at the center of the star. However, this possibility can only be checked numerically.

The constraints of Eq.~(\ref{con1}) and Eq.~(\ref{con2}) could give the expressions of $\sigma'$ and $a'$. Because Eq.~(\ref{con1}) contains the square of $\sigma'$, two branches are necessarily obtained,
\bea
\label{sc}
\sigma '(r) &=& \frac{{{e^{ - a}}}}{{{r^3}V}}\left( 2\left( {{d^2} - {r^2}{e^a}}  \right)V'
 \mp  F^{\frac{1}{2}} \right),\\
\label{ac}
a'(r) &=& \frac{{{e^{ - 2a}}\left( {{r^2}{e^a} - {d^2}} \right)V'}}{{{r^5}{V^2}}}\left( {2\left( {{r^2}{e^a} - {d^2}} \right)V' \pm {F^{\frac{1}{2}}}} \right) + \frac{{2\left( {{e^b} - 1} \right)}}{r},
\eea
where
\begin{align*}
\label{f}
F&=- 2{r^5}
{e^{a + 2b}}\left( {{e^a}(3r - 2d) - r} \right){V^3} - 2{r^4}{e^{2a}}\left( {{e^{2b}}\left( {{r^2}(8\pi p - W ) + 1} \right) - 4{e^b} + 3} \right){V^2} \\
&+ 4{{\left( {{d^2} - {r^2}{e^a}} \right)}^2}{{V'}^2}.
\end{align*}
They will be referred to as the negative or the positive branch respectively according to the sign in front of $F$ in Eq.~(\ref{sc}). It is obvious that the branching happening here is not physical and the two branches can be pieced together smoothly. In fact, two branches are equivalent under the transformation $\sigma \xrightarrow {} - \sigma$. We only keep the formalism in later calculations for convenience. 

It is necessary to emphasize that the system is not overdetermined, even though two independent equations of both $a$ and $\sigma$ are acquired. This fact can be understood by considering the hypothetical process of obtaining the TOV equation, in which all the field equations are reduced to a single one involving $p$ and $\rho$ except for the scalar. First, we could in principle rewrite $b$, $d$ and their derivatives in terms of $a$ and $\sigma$ because there are four equations at our disposal, namely (\ref{bc1}), (\ref{con1}), (\ref{con2}) and (\ref{bc2}). Then we could replace $b$, $d$ and their derivatives in Eq.~(\ref{a}) and Eq.~(\ref{s}). Finally, we could eliminate $a$ by using Eq.~(\ref{p}), arriving at two equations. One would be the TOV equation the other field equation for the scalar. There is no constraint left after the process and the system is not overdetermined as every equation is needed to get the final result. In short, Eq.~(\ref{con1}) and Eq.~(\ref{con2}) could in principle eliminate $b$ and $d$.

Some important conclusions can be drawn from Eq.~(\ref{sc}).
Suppose $\sigma=0$ at some $r$ which means $V'=0$. For relativistic star solutions, the term $e^{-a}$ is positive. If a non-vanishing graviton mass setup is considered then $V$ and $F^{\frac{1}{2}}$ are also positive which leads to the conclusion that  $\sigma$ could only approach $0$ from a negative value in the positive branch and from a positive value in the negative branch. A solution from one branch can cross $0$ and jump to another branch, forming an oscillating solution. In contrast, a non-oscillating solution must stay in one branch, such solutions are more desirable    \cite{Zhang:2017jze}.

In the case of vanishing graviton mass, the result is more complicated since a $0$-divided-by-$0$ situation may occur. Assuming that $\sigma$ approaches $0$ at some finite value of $r$, there is no reason to believe any component of the metric would be $0$ or divergent; they should each approach a finite value.  The denominator in $\sigma '(r)$ would be $\sim V_\sigma  r^3  \sigma^n $ and the first term in numerator reads $\sim 2n\left( {{d^2} - {r^2}{e^a}}  \right){V_\sigma}{\sigma ^{n-1}}$.  There are three terms approaching $0$ in $F$, the leading term is the last one which reads $\sim 4n^2{{\left( {{d^2} - {r^2}{e^a}} \right)}^2}{V_\sigma^2}{\sigma ^{2(n-1)}}$.
Depending on the sign of $ 2n\left( {{d^2} - {r^2}{e^a}}  \right){V_\sigma}{\sigma ^{n-1}}$, $\sigma'$ would approach $0$ or $\frac{{4n\left( {{d^2}{e^{-a}} - {r^2}} \right)}}{{{r^3}\sigma }}$.
$\sigma'$ approaching $0$ would cause $\sigma''$ to approach $0$, because in Eq.~(\ref{s}) it can be shown that if $\sigma = \sigma' =0$ then $\sigma'' =0$.  
Following the result, we can prove that all orders of derivatives of $\sigma$  equal to $0$ and $\sigma$ will stay at $0$ beyond that particular radius, leaving a reference metric component $d$ that is completely decoupled from the system.  This solution is mathematically possible but should be discarded because it is very difficult to comprehend physically. Not only does the scalar field remain $0$ beyond a finite radius while having a non-trivial structure near $r=0$, but also the degrees of freedom of the system change crossing the point of $\sigma=0$.

We are left with the possibility that $\sigma'$ approaches $\frac{{4n\left( {{d^2}{e^{-a}} - {r^2}} \right)}}{{{r^3}\sigma }}$ at $\sigma = 0$ thus diverges. To avoid such divergence, $\sigma$ should never cross $0$, which suggests that a non-trivial oscillating solution does not exist if $V$ is allowed to be $0$, the situation is the same in the full model. Previous studies in MVMG showed that oscillating solutions are quite normal \cite{Tolley:2015ywa,Zhang:2017jze,Sun:2019niy}. The lack of oscillating solutions may not be a problem theoretically, but they serve as useful tools to obtain the non-oscillating solution in numerical methods.

\section{The absence of well-behaved asymptotically flat solutions}
\label{sec:asymptotically flat solution}

The necessary condition for the existence of a physical relativistic star solution is the existence of a well-defined asymptotic solution. If such a solution is found it can also be used to integral inward to obtain a full solution. Unlike the previous studies on relativistic stars in MVMG, we find that with a diagonal metric the solutions in the cases of non-vanishing and vanishing graviton mass are quite different. Therefore, we have to distinguish between the two situations.

We can check whether a background is possible by plugging the corresponding metrics into the effective energy-momentum tensor $X^{\mu} _{ \nu}$. For more thorough discussions on background solutions, see \cite{Katsuragawa:2013lfa} and \cite{Katsuragawa:2013bma}. 
 
In the case of non-vanishing graviton mass, there is no solution for $d(r)$ to support a de Sitter or an anti-de Sitter background, which requires $X^{\mu} _{ \nu} \sim  \delta^{\mu} _{ \nu}$ at large $r$. In the minimal model, a possible workaround is that $d(r)$ approaches a constant which makes the reference  metric degenerate. A model with such a reference metric no longer belongs to dRGT. The existence of a flat background requires $X^{\mu} _{ \nu}$ to approach $0$ at large $r$, which implies  $d(r) \sim  r$ or any other slower growing function. However, the only solution that ensures a non-degenerate reference metric is $d(r) \sim  r$. This result holds in both the minimal model and the full model.

In the case of vanishing graviton mass, as the scalar approaches $0$, $V$ also decreases and the effect of $X^{\mu} _{ \nu}$ disappears. A flat background should be restored, while the requirement for its existence is more difficult to tell because $X^{\mu} _{ \nu}$ can have any value as long as  $V(\sigma) X_{\mu \nu}$ approaches $0$. In order to find the suitable flat background solution, we assume
$a(r)= a_0 + \delta a(r)$, $b(r)= b_0 + \delta b(r)$, $\sigma (r)= \delta \sigma(r)$ and $d(r) = d_{-1} (r) + \delta d(r)$ in which the $\delta$ terms represent infinitesimals and the rest of the functions or constants are finite. We plug them into the field equations of the full model and take the leading term, solving $d_{-1}(r)$ as
\be
d_{-1}(r) = \frac{e^{a_0}  (\alpha +\beta )-\beta  \pm \sqrt{ e^{2 a_0} \left(\alpha ^2-\beta \right)-e^{a_0} \left(2 \alpha ^2-\alpha -2 \beta -1\right)+\alpha ^2-\alpha -\beta }}{\left(e^{a_0}-1\right) \beta +\alpha -1}r.
\ee
This result is valid for both the full and minimal models, and we will use the assumption that $d (r) \sim  r$ in the case of vanishing graviton mass. Here we only focus on the solution of $d_{-1}(r)$ and ignore the rest of the perturbation equations; there will be cases in which the solution as a whole does not exist. We will discuss the issue in later sections.

In summary, we would focus on finding asymptotically flat solutions in all cases. Without loss of generality, the positive branch is chosen in the following calculations because the positive  and negative branches are equivalent under the transformation $\sigma \xrightarrow{}- \sigma$ in the current scalar potential setup. And we only focus on non-oscillating solutions which will stay in one branch. 

\subsection{Minimal Model}
We study the case of non-vanishing graviton mass in the minimal model by first considering the possibility that the asymptotically flat solutions are not analytic. Such an assumption stems from the observation that the solutions might contain a Yukawa term that is not analytic if $m \neq 0$. The solution should take the form of $a(r)= a_0 + \delta a(r)$, $b(r)= b_0 + \delta b(r)$, $\sigma (r)= \delta \sigma(r)$ and $d(r) = d_{-1}  r +d_0+ \delta d(r)$. As before the $\delta$ terms represent infinitesimals. A constant term $d_0$ is also included in $d(r)$ to represent the most general solution possible, as the constant term might be masked by the leading term in $d(r)$.
We set $n$, $l$ in Eq.~(\ref{vform}) and Eq.~(\ref{wform}) to $2$ and solve the linearized system of field equations to obtain the following result,
\bea
\delta a(r) &=& -\frac{\mathcal{C}_0  e^{-\sqrt{m} r}}{\sqrt{m} r},\n\\
\delta b(r) &=& \frac{\mathcal{C}_0 e^{-\sqrt{m} r} \left(\sqrt{m} r+1\right)}{2 \sqrt{m} r},\n\\
\delta d(r) &=& -\frac{\mathcal{C}_0 e^{-\sqrt{m} r} \left(m r^2+\sqrt{m} r+1\right)}{2 m^{3/2} r^2},\n\\
\label{solY}
\delta \sigma(r) &=& \frac{\mathcal{C}_1 e^{-r \sqrt{2W_{\sigma}}}}{r},
\eea
also $a_0 = b_0 = d_0 = 0$ and $d_{-1}=1$.
What is peculiar about this solution is that it lacks a Schwarzschild background, resulting in an arbitrarily small ADM mass and its inability to describe a physical star. The role of constraints Eq.~(\ref{sc}) and Eq.~(\ref{ac}) in obtaining the solution is also noteworthy. We first solve the linearized system of field equations without the two constraints, the solution is a combination of the Schwarzschild background and a Yukawa term. We then plug the solution into the linearized constraints and find that the Schwarzschild background must be absent for the constraints to hold, which can be achieved by setting the constant factor of the $1/r$ term in the Schwarzschild background to $0$. To a certain degree, the existence of the constraints makes the physical stars impossible in the context of non-analytic solutions.

To check whether the results hold in more general forms of potential, we raise the value of $n$, $l$ to $4$. We find that the non-analytic, asymptotically flat solution still exists and it is the same as Eq.~(\ref{solY}) except for $\delta \sigma(r)$ which reads
\be
\delta \sigma(r)=-\frac{\mathcal{C}_1}{r}.
\ee
For $n$ and $l$ that are bigger than $4$, the same conclusion holds as the higher order derivatives of the potentials do not enter the linearized system of field equations.

The asymptotic solution could also be analytic, we explore the possibility by setting
\bea
a\left( r \right) &=& \sum\limits_{i = 0} {{a_i}\frac{1}{{{r^i}}}}, \n\\
b\left( r \right) &=& \sum\limits_{i = 0} {{b_i}\frac{1}{{{r^i}}}}, \n\\
\sigma\left( r \right) &=& \sum\limits_{i = 1} {{\sigma_i}\frac{1}{{{r^i}}}}, \n\\
d(r) &=&  d_{-1} r + d_0 + \sum\limits_{i = 1} {{d_{i}} \frac{1}{{{r^i}}}}.\label{apsol}
\eea
We plug the series Eq.~(\ref{apsol}) into the field equations and can only obtain a solution in which
\be
\sigma_1 = \sigma_2 = \sigma_3 = \cdots = 0. 
\ee
The value of $\sigma_1$ can be solved directly to be $0$, and the rest of $\sigma_i$ can be proven to be $0$ by expanding Eq.~(\ref{s}). The expression of each order is a polynomial of $\sigma_i$ with non-zero coefficients such that the condition  $\sigma_1 = 0$ is sufficient to conclude $\sigma_i = 0$ for higher values of $i$. The solution will lead to a constant scalar field and consequently to a constant graviton mass. Therefore, this relativistic star solution would  reduce to that of dRGT and should be considered trivial under MVMG. We increase the value of $n$, $l$ to $4$ and can only find the same trivial solution. 

At this point, we can justify the uniqueness of the asymptotic solution in the case of non-vanishing graviton mass, which is related to the discussions in the previous section. If the minimal model is considered, in the case of a non-analytic solution, the condition of uniqueness is obvious as we only obtain one solution. In the case of an analytic solution, the trivial solution is the same as the one obtained in minimal dRGT with a diagonal metric, which does not have branching \cite{Katsuragawa:2015lbl}. In the case of vanishing graviton mass, the condition of uniqueness may not hold, and we will discuss this subject in the rest of the section.

We set $m=0$ to include the case of vanishing graviton mass. Using the same method of solving the linearized system of field equations, we aim to obtain possible non-analytic, asymptotically flat solutions. We find that for $n = l = 2$ there is no suitable solution at all, as the solution of $\delta d(r)$ is not an infinitesimal at large $r$. The situation is similar if $n$ and $l$ are increased to 4, in which case the
solution of  $\delta d(r)$ is a constant associated with $\delta a(r)$ and $\delta b(r)$ which cannot be set to $0$.

We try to find analytic solutions next by assuming the same approximate solution as in Eq.~(\ref{apsol}).
However, the choice of $d(r)$ deserves some clarification because the approach to obtaining the conclusion $d(r)\sim r$ in the case of vanishing graviton mass could not cover the case of analytic solutions. We can study the possibility that $d(r)$ contains higher order of $r$ by instead assuming  
\be
d(r) =d_{-2} r^2 +   d_{-1} r + d_0 + \sum\limits_{i = 1} {{d_{i}} \frac{1}{{{r^i}}}},
\ee
and solve the field equations. It is straightforward to show that $d_{-2} = 0$. It can also be proven in the same manner that higher order terms are $0$ too.
As such, we can conclude that $d(r)\sim r$ in the case of analytic solutions. After some calculations, we find that for $n = l = 2$, only a trivial solution with a constant scalar exists. The reason is similar to that of the case of non-vanishing graviton mass. But if we increase $n$ and $l$ to 4, there seems to be a non-trivial solution besides the trivial one that reads
\bea
b_0 &=& 0, \n\\
a_0 &=& \ln \left(d_{-1}^2\right), \n\\
d_0 &=& \frac{3 a_1 d_{-1}}{8}.
\eea
Also $d_{-1}$ is determined by 
\be
2 V^{(4)} d_{-1}^3 + (W^{(4)}-3 V^{(4)}) d_{-1}^2 +V^{(4)} =0.
\ee
The derivatives of the potentials are taken at $\sigma =0$.
The remaining coefficients can be solved order by order for given $a_1$ and $\sigma_1$. In other words, $a_1$ and $\sigma_1$ are free to choose and there are three branches corresponding to one such choice. This solution can potentially describe a physical star, but the problem discussed in Section \ref{sec:branching} will arise since it has branches, namely a mismatch between the parameter space of the solution at $r = 0$ and that of infinity. The validity of the solution can only be checked by numerical methods, we will investigate this solution further in future studies.

\subsection{Full Model}
\label{sec:general}
In this section, we study the relativistic star solutions in the full model with the same settings for the mass and scalar potentials as in the minimal model. The structure of the system is still the same,  DAEs of index-2 with two constraints. However, the field equations become too lengthy and inconvenient to write down, especially for $b$, $d$ and the two constraints. 

We first study the non-analytic, asymptotically flat solutions in the case of non-vanishing graviton mass by setting $a(r)= a_0 + \delta a(r)$, $b(r)= b_0 + \delta b(r)$, $\sigma (r)= \delta \sigma(r)$ and $d(r) = d_{-1}  r + d_0+ \delta d(r)$. In the full model, there will be more background solutions that satisfy the field equation compared to the minimal model. After solving the linearized system of field equations, we find that for any value of $n$ and $l$, the solutions in the minimal model are reproduced, there also exists a new solution in a very similar form with a different background, which reads
\bea
\label{a_0}
a_0 = \ln{\frac{\beta  d_{-1}^2 -2  (\alpha +\beta )d_{-1}+2 \alpha +\beta+1}{(\alpha +\beta )d_{-1}^2 -2 (2 \alpha +\beta +1)d_{-1} +3 \alpha +\beta +3}}.
\eea
The solution for $d_{-1}$ is multi-valued,
\be
\label{d_-1}
d_{-1}=\frac{3 \alpha +2 \beta \pm  \sqrt{3 \left(3 \alpha ^2-4 \beta \right)}}{2 \beta }.
\ee
Also, we have $b_0=0$. This solution is essentially the same as Eq.~(\ref{solY}) but with different constants, so it also has an arbitrarily small ADM mass and cannot describe a physical star.
A new and different solution also exists that gives the following background, but it requires  $\beta = \frac{3 }{4}\alpha ^2$.
\bea
a_0 &=& \ln \left(\frac{\alpha }{\alpha +2}\right),\n\\
b_0 &=& 0, \n\\
d_{-1} &=& \frac{\alpha +2}{\alpha },
\eea
it is in agreement with Eq.~(\ref{a_0}) and Eq.~(\ref{d_-1}). The unique feature of this solution is that it is analytic by not possessing a Yukawa term even with a finite graviton mass. However, as an analytic solution, its validity should be checked accordingly. Simply linearizing the system does not properly reflect the magnitude relation between $1/r$ and the infinitesimals in the approximate solution at spatial infinity. 

We move on to the analytic asymptotically flat solution by expanding the system in terms of $1/r$ with the same settings as in the minimal model. In all of the situations we could only recover trivial solutions, even if $\beta = \frac{3 }{4}\alpha ^2$. Still, that particular choice of parameters might hold some interesting properties in dRGT. 

In the case of vanishing graviton mass, we first study the non-analytic solution. For the same reason as in the minimal model, no valid solution exists for $\delta d(r)$ in both cases of $n=l=2$ and $n=l=4$.
As for analytic solutions, we could only recover a trivial solution for $n=l=2$. However, if we set $n=l=4$ there is a non-trivial solution in addition to the trivial one, just as in the minimal model.
It is characterized by the value of $d_{-1}$, given by the following quartic equation,
\bea
& & V^{(4)}(\alpha^{2} - \alpha -\beta ) d_{-1}^4 + 2 V^{(4)} \left(-2 \alpha ^2+\alpha +2 \beta +1\right) d_{-1}^3 \n\\
& &+\left( V^{(4)} \left(6 \alpha ^2-6 \beta -3\right)+W^{(4)} (-\alpha +\beta +1) \right)d_{-1}^2 
-2 \left(V^{(4)} \left(2 \alpha ^2+\alpha -2 \beta \right)+\beta  W^{(4)}\right) d_{-1}\n\\
& &+ V^{(4)} \left(\alpha ^2+\alpha -\beta +1\right)+W^{(4)} (\alpha +\beta ) = 0.
\eea
All the derivatives of the potentials are taken at $\sigma =0$. We also have
\be
a_0 =\ln{\left(\frac{V^{(4)} \left(  (\alpha -\beta -1)d_{-1}^3+3 \beta  d_{-1}^2 -3  (\alpha +\beta  ) d_{-1} +1 + 2 \alpha +\beta \right)}{d_{-1} V^{(4)} \left( -\beta  d_{-1}^2 +3  (\alpha +\beta ) d_{-1} -3 (2 \alpha +\beta +1)\right)+V^{(4)} (3 \alpha +\beta +3)-W^{(4)}}\right)}, 
\ee
and $b_0 = 0$. The rest of the coefficients can be solved order by order except for $a_1$ and $\sigma_1$.  The solution is analogous to the one in the minimal model with the same problem and a much more complicated equation on $d_{-1}$. If we increase $n$ and $l$ to higher values, the solution still has multiple branches but the equation regarding the different branches changes yet again.  If one could draw a conclusion, it is that for vanishing graviton mass one possible analytic solution with multiple branches exists, but its form is dependent on potentials, in particular the mass potential $V$.

\section{Summary and discussion}
\label{sec:summary}
We have studied the relativistic star solutions in MVMG with a diagonal physical metric. The system is dramatically different from the one with a non-diagonal physical metric \cite{Sun:2019niy}, in the sense that field equations form a system of  DAEs of index-2 instead of a system of ODEs. We transform the system by taking derivatives of the field equation that function as an algebraic constraint twice, and obtain a system of ODEs with two additional constraints.

The first observation that can be made about the system is that there is no possibility of an oscillating solution in the case of vanishing graviton mass, due to the two additional constraints. The absence of oscillating solutions is unusual in MVMG, as they can be found in the previous studies \cite{Tolley:2015ywa,Zhang:2017jze,Sun:2019niy}. It may also lead to difficulties in obtaining non-oscillating solutions because such solutions are usually the outcome of fine-tuning the oscillating solutions.

We focus the rest of the paper on finding the approximations of the possible solutions at a large radius. In the case of a diagonal physical metric, we find that the choice of mass potential and scalar potential affects the solutions of the system at a fundamental level, which is very different from the case of a non-diagonal physical metric.
To include as many circumstances as possible, we study $4$ cases separately based on whether graviton mass is non-vanishing or not and whether mass potential and scalar potential are quadratic or higher.

Overall, there are three types of asymptotically flat solutions. In both the minimal model and the full model, trivial solutions can be found, which have a constant graviton mass and are not meaningful in MVMG. In the case of non-vanishing graviton mass, there exist non-analytic solutions without the Schwarzschild background, caused directly by the two constraints. The solutions are not suitable to describe physical stars because the ADM mass of the solutions is $0$. Finally, we have found noteworthy solutions if the graviton mass is allowed to be $0$, and the mass potential  is quartic or higher. These solutions take the standard form of approximate solutions and can be expressed in a series of $1/r$. They might be candidates for describing physical stars,  however, these solutions suffer from the problem of having multiple branches. Because the approximate solution at the center of the star is unique,  it cannot be matched one-to-one with the asymptotic solution and may lead to other issues. We will investigate this type of solution further in future studies.

In the case of non-vanishing graviton mass, we can conclude that there is no possible non-trivial physical star solution. In the case of vanishing graviton mass, there are possible candidates for physical star solutions if their unusual nature is ignored, however, the oscillating solutions are missing which makes the task of finding possible solutions numerically very difficult, as an improper choice of parameters will always lead to a divergent solution instead of an oscillating one.
We tried to find relativistic star solutions numerically using boundary conditions generated by approximate solutions at the center of the star. However, we have not been able to find suitable solutions as they tend to diverge very fast.  
As a result, a non-diagonal physical metric is much preferable when studying relativistic star solutions in MVMG.  
The findings in this paper might be related to the difficulties in the studies of relativistic star solutions in dRGT with a diagonal physical metric \cite{Yamazaki:2018szv}, in which case a non-diagonal physical metric might be worth considering. 
It seems that by introducing more degrees of freedom, the extra constraints leading to the absence of well-behaved solutions will be eliminated, which could be accomplished by introducing a non-diagonal physical metric or a dynamical reference metric.

\section*{Acknowledgement}
We would like to thank Shuang-Yong Zhou, Xue Sun and Jun Zhang for helpful suggestions and discussions. DJW is supported by NSFC, No. 11947104.


\begin{thebibliography}{99}

\bibitem{deRham:2010ik}
  C.~de Rham, G.~Gabadadze,
  Phys.\ Rev.\  {\bf D82}, 044020 (2010),
  [arXiv:1007.0443].

\bibitem{deRham:2010kj}
  C.~de Rham, G.~Gabadadze and A.~J.~Tolley,
  Phys.\ Rev.\ Lett.\  {\bf 106}, 231101 (2011),
  [arXiv:1011.1232].


\bibitem{Hassan:2011hr}
  S.~F.~Hassan and R.~A.~Rosen,
  Phys.\ Rev.\ Lett.\  {\bf 108}, 041101 (2012)
  [arXiv:1106.3344 [hep-th]].

\bibitem{Hassan:2011tf}
 S.~F.~Hassan, R.~A.~Rosen and A.~Schmidt-May,
 JHEP \textbf{02}, 026 (2012)
 doi:10.1007/JHEP02(2012)026
 [arXiv:1109.3230 [hep-th]].

\bibitem{Hassan:2011ea}
  S.~F.~Hassan and R.~A.~Rosen,
  JHEP {\bf 1204} (2012) 123
  [arXiv:1111.2070 [hep-th]].

\bibitem{deRham:2011rn}
 C.~de Rham, G.~Gabadadze and A.~J.~Tolley,
 Phys. Lett. B \textbf{711}, 190-195 (2012)
 doi:10.1016/j.physletb.2012.03.081
 [arXiv:1107.3820 [hep-th]].

\bibitem{deRham:2014zqa}
  C.~de Rham,
  Living Rev.\ Rel.\  {\bf 17}, 7 (2014)
  [arXiv:1401.4173 [hep-th]].
  
\bibitem{Hinterbichler:2011tt}
  K.~Hinterbichler,
  Rev.\ Mod.\ Phys.\  {\bf 84}, 671 (2012)
  [arXiv:1105.3735 [hep-th]].  


\bibitem{Hassan:2011zd}
S.~F.~Hassan and R.~A.~Rosen,
JHEP \textbf{02}, 126 (2012)
doi:10.1007/JHEP02(2012)126
[arXiv:1109.3515 [hep-th]].

\bibitem{DAmico:2012hia}
G.~D'Amico, G.~Gabadadze, L.~Hui and D.~Pirtskhalava,
Phys. Rev. D \textbf{87}, 064037 (2013)
doi:10.1103/PhysRevD.87.064037
[arXiv:1206.4253 [hep-th]].

\bibitem{DeFelice:2013tsa}
A.~De Felice and S.~Mukohyama,
Phys. Lett. B \textbf{728}, 622-625 (2014)
doi:10.1016/j.physletb.2013.12.041
[arXiv:1306.5502 [hep-th]].

\bibitem{Huang:2012pe}
Q.~G.~Huang, Y.~S.~Piao and S.~Y.~Zhou,
Phys. Rev. D \textbf{86}, 124014 (2012)
doi:10.1103/PhysRevD.86.124014
[arXiv:1206.5678 [hep-th]].

\bibitem{Huang:2013mha}
Q.~G.~Huang, K.~C.~Zhang and S.~Y.~Zhou,
JCAP \textbf{08}, 050 (2013)
doi:10.1088/1475-7516/2013/08/050
[arXiv:1306.4740 [hep-th]].

\bibitem{Zhang:2017jze}
J.~Zhang and S.~Y.~Zhou,
Phys. Rev. D \textbf{97}, no.8, 081501 (2018)
doi:10.1103/PhysRevD.97.081501
[arXiv:1709.07503 [gr-qc]].

\bibitem{Riess:1998cb}
A.~G.~Riess \textit{et al.} [Supernova Search Team],
Astron. J. \textbf{116}, 1009-1038 (1998)
doi:10.1086/300499
[arXiv:astro-ph/9805201 [astro-ph]].


\bibitem{Perlmutter:1998np}
S.~Perlmutter \textit{et al.} [Supernova Cosmology Project],
Astrophys. J. \textbf{517}, 565-586 (1999)
doi:10.1086/307221
[arXiv:astro-ph/9812133 [astro-ph]].

\bibitem{Clifton:2011jh}
T.~Clifton, P.~G.~Ferreira, A.~Padilla and C.~Skordis,
Phys. Rept. \textbf{513}, 1-189 (2012)
doi:10.1016/j.physrep.2012.01.001
[arXiv:1106.2476 [astro-ph.CO]].

\bibitem{Abbott:2016blz}
B.~P.~Abbott \textit{et al.} [LIGO Scientific and Virgo],
Phys. Rev. Lett. \textbf{116}, no.6, 061102 (2016)
doi:10.1103/PhysRevLett.116.061102
[arXiv:1602.03837 [gr-qc]].

\bibitem{TheLIGOScientific:2017qsa}
B.~P.~Abbott \textit{et al.} [LIGO Scientific and Virgo],
Phys. Rev. Lett. \textbf{119}, no.16, 161101 (2017)
doi:10.1103/PhysRevLett.119.161101
[arXiv:1710.05832 [gr-qc]].

\bibitem{deRham:2016nuf}
C.~de Rham, J.~T.~Deskins, A.~J.~Tolley and S.~Y.~Zhou,
Rev. Mod. Phys. \textbf{89}, no.2, 025004 (2017)
doi:10.1103/RevModPhys.89.025004
[arXiv:1606.08462 [astro-ph.CO]].

\bibitem{Berezhiani:2011mt}
L.~Berezhiani, G.~Chkareuli, C.~de Rham, G.~Gabadadze and A.~J.~Tolley,
Phys. Rev. D \textbf{85} (2012), 044024
doi:10.1103/PhysRevD.85.044024
[arXiv:1111.3613 [hep-th]].

\bibitem{Babichev:2015xha}
E.~Babichev and R.~Brito,
Class. Quant. Grav. \textbf{32}, 154001 (2015)
doi:10.1088/0264-9381/32/15/154001
[arXiv:1503.07529 [gr-qc]].

\bibitem{Tolley:2015ywa}
A.~J.~Tolley, D.~J.~Wu and S.~Y.~Zhou,
Phys. Rev. D \textbf{92}, no.12, 124063 (2015)
doi:10.1103/PhysRevD.92.124063
[arXiv:1510.05208 [hep-th]].

\bibitem{Sun:2019niy}
X.~Sun and S.~Y.~Zhou,
Phys. Rev. D \textbf{101}, no.4, 044060 (2020)
doi:10.1103/PhysRevD.101.044060
[arXiv:1912.00685 [gr-qc]].

\bibitem{Katsuragawa:2015lbl}
T.~Katsuragawa, S.~Nojiri, S.~D.~Odintsov and M.~Yamazaki,
Phys. Rev. D \textbf{93}, 124013 (2016)
doi:10.1103/PhysRevD.93.124013
[arXiv:1512.00660 [gr-qc]].

\bibitem{Yamazaki:2017ztk}
M.~Yamazaki, T.~Katsuragawa, S.~Nojiri and S.~D.~Odintsov,
PoS \textbf{KMI2017}, 041 (2017)
doi:10.22323/1.294.0041

\bibitem{Yamazaki:2018szv}
M.~Yamazaki, T.~Katsuragawa, S.~D.~Odintsov and S.~Nojiri,
Phys. Rev. D \textbf{100}, no.8, 084060 (2019)
doi:10.1103/PhysRevD.100.084060
[arXiv:1812.10239 [gr-qc]].

\bibitem{Hendi:2017ibm}
S.~H.~Hendi, G.~H.~Bordbar, B.~Eslam Panah and S.~Panahiyan,
JCAP \textbf{07}, 004 (2017)
doi:10.1088/1475-7516/2017/07/004
[arXiv:1701.01039 [gr-qc]].

\bibitem{EslamPanah:2018evk}
B.~Eslam Panah and H.~L.~Liu,
Phys. Rev. D \textbf{99}, no.10, 104074 (2019)
doi:10.1103/PhysRevD.99.104074
[arXiv:1805.10650 [gr-qc]].


\bibitem{Katsuragawa:2013bma}
T.~Katsuragawa and S.~Nojiri,
Phys. Rev. D \textbf{87} (2013) no.10, 104032
doi:10.1103/PhysRevD.87.104032
[arXiv:1304.3181 [hep-th]].

\bibitem{Katsuragawa:2013lfa}
T.~Katsuragawa,
Phys. Rev. D \textbf{89} (2014), 124007
doi:10.1103/PhysRevD.89.124007
[arXiv:1312.1550 [hep-th]].

\end{thebibliography}
\end{document}